\documentclass{llncs}

\usepackage{amsmath, amsfonts}
\usepackage[colorinlistoftodos, textwidth=3.5cm]{todonotes}

\newtheorem{lemm}{Lemma}
\newtheorem{theo}[lemm]{Theorem}
\newtheorem{prop}[lemm]{Proposition}
\newtheorem{defi}[lemm]{Definition}

\newcommand{\Z}{{\mathbb{Z}}}

\newcommand{\dd}{\displaystyle}
\newcommand{\zz}{\Z_2\Z_4}
\newcommand{\codi}{{\cal C}}

\newcommand{\G}{{\mathcal{G}}}

\newcommand{\ARM}{{\mathcal{ARM}}}

\newcommand{\EQ}{\begin{equation}}
\newcommand{\EN}{\end{equation}}

\begin{document}

\title{Construction of Additive Reed-Muller Codes\thanks{This work has been partially supported by the Spanish MICINN Grants MTM2006-03250, TSI2006-14005-C02-01, PCI2006-A7-0616 and also by the \textit{Comissionat per a Universitats i Recerca de la Generalitat de Catalunya} under grant FI2008. Part of the material in this paper was presented without proofs at the $18$th Symposium on Applied algebra, Algebraic algorithms, and Error Correcting Codes (AAECC 2009), Tarragona, Spain, June 8-12, 2009.}
}

\author{J.~Pujol\inst{} \and J.~Rif\`{a}\inst{} \and L.~Ronquillo\inst{}}

\institute{Department of Information and Communications Engineering,
Universitat Aut\`{o}noma de Barcelona, 08193-Bellaterra, Spain.}

 \maketitle
 \thispagestyle{empty}

\begin{abstract}
The well known Plotkin construction is, in the current paper, generalized and used to yield new families of $\zz$-additive codes, whose length, dimension as well as minimum distance are studied. These new constructions enable us to obtain families of $\zz$-additive codes such that, under the Gray map, the corresponding binary codes have the same parameters and properties as the usual binary linear Reed-Muller codes. Moreover, the first family is the usual binary linear Reed-Muller family.
\bigskip

\end{abstract}

\section{Introduction}

The aim of our  paper is to obtain a generalization of the Plotkin construction which gave rise to families of $\zz$-additive codes such that, after the Gray map, the corresponding $\zz$-linear codes had the same parameters and properties as the family of binary linear $RM$ codes. Even more, we want the corresponding codes with parameters $(r,m)=(1,m)$ and $(r,m)=(m-2,m)$ to be, respectively, any one of the non-equivalent $\zz$-linear Hadamard and $\zz$-linear $1$-perfect codes.

\section{Constructions of $\zz$-additive codes}\label{sec2}

In general, any non-empty subgroup ${\cal C}$ of $\Z_2^\alpha \times
\Z_4^\beta$ is a {\it $\zz$-additive code}, where $\Z_2^\alpha$ denotes the set of all binary
vectors of length $\alpha$ and $\Z_4^\beta$ is the set of all $\beta$-tuples in $\Z_4$.

Let $\codi$ be a $\zz$-additive code, and let $C=\Phi(\codi)$, where
$\Phi: \Z_2^{\alpha}\times\Z_4^{\beta} \longrightarrow \Z_2^{n}$ is given by the map $\Phi(u_1, \ldots, u_{\alpha} | v_1, \ldots, v_{\beta})=(u_1, \ldots, u_{\alpha} | \phi(v_1), \ldots, \phi(v_{\beta}))$ where $\phi(0)=(0,0)$, $\phi(1)=(0,1)$, $\phi(2)=(1,1)$, and $\phi(3)=(1,0)$ is the usual Gray map from $\Z_4$ onto $\Z_2^2$.

Since the Gray map is distance preserving, the Hamming distance of a $\zz$-linear code $C$ coincides with the Lee distance computed on the $\zz$-additive code $\codi=\phi^{-1}(C)$.

A $\zz$-additive code $\codi$ is also isomorphic to an abelian structure like
$\Z_2^{\gamma}\times \Z_4^{\delta}$. Therefore, $\codi$ has $|\codi|=2^\gamma 4^\delta $ codewords and, moreover, $2^{\gamma+\delta}$ of them are of order two. We call such code $\codi$ a
{\it $\zz$-additive code of type $(\alpha,\beta;\gamma,\delta)$} and its
binary image $C=\Phi(\codi)$ is a {\it $\zz$-linear code of type
$(\alpha,\beta;\gamma,\delta)$}. 

Although $\codi$ may not have a basis, it is important and
appropriate to define a generator matrix for $\codi$ as:
\EQ\label{generatorMatrix}
   \G=\left ( \begin{array}{c|c}
        B_2 & Q_2 \\
        \hline
        B_4 & Q_4 \\
    \end{array}\right ),
\EN
where $B_2$ and $B_4$ are binary matrices of size $\gamma\times
\alpha$ and $\delta\times \alpha$, respectively; $Q_2$ is a $\gamma\times \beta$-quaternary matrix which contains order two row vectors; and $Q_4$ is a $\delta\times \beta$-quaternary matrix with order four row vectors.

\subsection{Plotkin construction}

In this section we show that the well known Plotkin construction can be generalized to $\zz$-additive codes.

\begin{defi}[Plotkin Construction]\label{PlotDefi}
 Let
$\mathcal{X}$ and $\mathcal{Y}$
be any two $\zz$-additive codes of types
$(\alpha,\beta;\gamma_\mathcal{X},\delta_\mathcal{X})$,
$(\alpha,\beta;\gamma_\mathcal{Y},\delta_\mathcal{Y})$ and minimum distances
$d_\mathcal{X}$, $d_\mathcal{Y}$, respectively. If
$\G_\mathcal{X}$ and $\G_\mathcal{Y}$ are the
generator matrices of $\mathcal{X}$ and $\mathcal{Y}$, then the
matrix
$$
\G_{P}=\left ( \begin{array}{cc}
                         \G_\mathcal{X} & \G_\mathcal{X} \\
                          0  & \G_\mathcal{Y} \\
                      \end{array} \right )
$$ is the generator matrix of a new $\zz$-additive code $\mathcal{C}$.
\end{defi}

\begin{prop}\label{PlotConst}
Code $\mathcal{C}$ defined above is a
$\zz$-additive code of type $(2\alpha,2\beta;\gamma, \delta)$, where
$\gamma=\gamma_\mathcal{X}+\gamma_\mathcal{Y}$,
$\delta=\delta_\mathcal{X}+\delta_\mathcal{Y}$, binary length $n=2\alpha+4\beta$, size $2^{\gamma+2\delta}$ and minimum distance
$d=\min\{2d_\mathcal{X},d_\mathcal{Y}\}$.
\end{prop}

\subsection{BA-Plotkin construction}

Given a $\zz$-additive code $\mathcal{C}$ with generator matrix $\G$ we denote,
respectively, by $\G[b_2]$, $\G[q_2]$, $\G[b_4]$ and $\G[q_4]$ the four
submatrices $B_2$, $Q_2$, $B_4$, $Q_4$ of $\G$ defined in
(\ref{generatorMatrix}); by $\G[b]$ and $\G[q]$ the submatrices $\left (
\begin{array}{c|}
        B_2 \\
        \hline
        B_4 \\
    \end{array}\right )$, $\left ( \begin{array}{|c}
        Q_2 \\
        \hline
        Q_4 \\
    \end{array}\right )$; and by $\G[bq_2]$ and $\G[bq_4]$ the submatrices $\left ( \begin{array}{c|c}
        B_2 & Q_2 \\
\hline
    \end{array}\right )$, $\left ( \begin{array}{c|c}
        \hline
B_4 & Q_4
    \end{array}\right )$, respectively.

\begin{defi}[Double-Plotkin Construction]
 Let $\mathcal{X}$, $\mathcal{Y}$ and $\mathcal{Z}$ be three $\zz$-additive codes with generator matrices $\G_\mathcal{X}$, $\G_\mathcal{Y}$ and $\G_\mathcal{Z}$, respectively. By applying two Plotkin constructions, one after another, we obtain a $\zz$-additive code with generator matrix

$$
\left ( \begin{array}{c c c c | c c c c}
        \G_\mathcal{X}[b] & \G_\mathcal{X}[b] & \G_\mathcal{X}[b] & \G_\mathcal{X}[b] & \G_\mathcal{X}[q] & \G_\mathcal{X}[q] & \G_\mathcal{X}[q] & \G_\mathcal{X}[q] \\
        \G_\mathcal{Y}[b] & \G_\mathcal{Y}[b] & 0 & 0 & 0 & 2\G_\mathcal{Y}[q] & \G_\mathcal{Y}[q] & 3\G_\mathcal{Y}[q] \\
	0 & \G_\mathcal{Y}[b] & 0 & \G_\mathcal{Y}[b] & 0 & \G_\mathcal{Y}[q] & 2\G_\mathcal{Y}[q] & 3\G_\mathcal{Y}[q] \\
	\G_\mathcal{Y}[b] & \G_\mathcal{Y}[b] & 0 & 0 & 0 & 0 & \G_\mathcal{Y}[q] & \G_\mathcal{Y}[q] \\
	0 & \G_\mathcal{Z}[b] & 0 & 0 & 0 & 0 & 0 & \G_\mathcal{Z}[q] \\
    \end{array}\right ).
$$

\end{defi}

\noindent We can see the above matrix as having $5$ row submatrices and $8$ column submatrices.
By slightly changing these submatrices, we obtain a new
construction with interesting properties with regard to the minimum
distance of the generated code. We call this new construction {\it
BA-Plotkin construction}.

\medskip

%%%%%%%%%%%%%%%%%%%%

\begin{defi}[BA-Plotkin Construction] \label{BAPlot}
Let $\mathcal{X}$, $\mathcal{Y}$ and $\mathcal{Z}$ be any three
$\zz$-additive codes of types
$(\alpha,\beta;\gamma_\mathcal{X},\delta_\mathcal{X})$,
$(\alpha,\beta;\gamma_\mathcal{Y},\delta_\mathcal{Y})$,
$(\alpha,\beta;\gamma_\mathcal{Z},\delta_\mathcal{Z})$, generator matrices $\G_\mathcal{X}$, $\G_\mathcal{Y}$ and
$\G_\mathcal{Z}$, and minimum distances
$d_\mathcal{X}$, $d_\mathcal{Y}$, $d_\mathcal{Z}$, respectively; such that $\G_\mathcal{Z}$ $\subset$ $\G_\mathcal{Y}$ $\subset$ $\G_\mathcal{X}$ and $d_\mathcal{Z}=2d_\mathcal{Y}=4d_\mathcal{X}$.

\noindent We define a new $\zz$-additive code $\mathcal{C}$ with generator matrix
$$
\G_{BA} = \left ( \begin{array}{c c | c c c c c}
\G_\mathcal{X}[b] & \G_\mathcal{X}[b] & 2\G_\mathcal{X}[b] & \G_\mathcal{X}[q] & \G_\mathcal{X}[q] & \G_\mathcal{X}[q] & \G_\mathcal{X}[q]\\
0 & \G_\mathcal{Y}[b_2] & \G_\mathcal{Y}[b_2] & 0 & 2\G'_\mathcal{Y}[q_2] & \G'_\mathcal{Y}[q_2] & 3\G'_\mathcal{Y}[q_2]\\
0 & \G_\mathcal{Y}[b_4] & \G_\mathcal{Y}[b_4] & 0 & \G_\mathcal{Y}[q_4] & 2\G_\mathcal{Y}[q_4] & 3\G_\mathcal{Y}[q_4]\\
\G_\mathcal{Y}[b_4] & \G_\mathcal{Y}[b_4] & 0 & 0 & 0 & \G_\mathcal{Y}[q_4] & \G_\mathcal{Y}[q_4]\\
0 & \G_\mathcal{Z}[b] & 0 & 0 & 0 & 0 & \G_\mathcal{Z}[q]
\end{array} \right ),
$$
where $\G'_\mathcal{Y}[q_2]$ is the matrix obtained from $\G_\mathcal{Y}[q_2]$
after switching twos by ones in its $\gamma_\mathcal{Y}$ rows of order two, and considering the ones from the third column submatrix of the construction as ones in the quaternary ring $\Z_4$. 

\end{defi}

\begin{prop}\label{BA-PlotkinConst}
Code $\mathcal{C}$ generated by the BA-Plotkin
construction from Definition~\ref{BAPlot} is a $\zz$-additive code of minimum Lee distance $d = 4d_\mathcal{X}$.
\end{prop}

\begin{proof}

Let $\mathcal{X}$, $\mathcal{Y}$ and $\mathcal{Z}$ be any three $\zz$-additive
codes with generator matrices $\G_\mathcal{X}$, $\G_\mathcal{Y}$ and
$\G_\mathcal{Z}$, respectively, such that $\G_\mathcal{Z}$ $\subset$
$\G_\mathcal{Y}$ $\subset$ $\G_\mathcal{X}$ and
$d_\mathcal{Z}=2d_\mathcal{Y}=4d_\mathcal{X}$. Let $\mathcal{Y}[b_2]$,
$\mathcal{Y}'[q_2]$,$\mathcal{Y}[bq_2]$ and $\mathcal{Y}[bq_4]$ be the codes
generated by $\G_\mathcal{Y}[b_2]$, $\G'_\mathcal{Y}[q_2]$,
$\G_\mathcal{Y}[bq_2]$ and $\G_\mathcal{Y}[bq_4]$. Let $d_{\mathcal{Y}[bq_2]}
\geq d_{\mathcal{Y}}$ and $d_{\mathcal{Y}[bq_4]} \geq d_{\mathcal{Y}}$ be the
minimum Lee distances of codes $\mathcal{Y}[bq_2]$ and $\mathcal{Y}[bq_4]$,
respectively.

\bigskip

Let $\textbf{u}_1$ be any vector in $\mathcal{X}$, $\textbf{u}_2$ $\in$
($\mathcal{Y}[b_2],\mathcal{Y}'[q_2]$), $\textbf{u}_3$, $\textbf{u}_4$ $\in$
$\mathcal{Y}[bq_4]$ and $\textbf{u}_5$ $\in$ $\mathcal{Z}$. Note that
$(\mathcal{Y}[b_2],2\mathcal{Y}'[q_2]) = \mathcal{Y}[bq_2]$.

Let $\textbf{f}_1=(\textbf{u}_1[b],\textbf{u}_1[b]|2\textbf{u}_1[b],\textbf{u}_1[q],\textbf{u}_1[q],\textbf{u}_1[q],\textbf{u}_1[q])$,\\ 
$\textbf{f}_2=(0, \textbf{u}_2[b] | \textbf{u}_2[b], 0, 2\textbf{u}_2[q], \textbf{u}_2[q], 3\textbf{u}_2[q])$,
$\textbf{f}_3=(0, \textbf{u}_3[b] | \textbf{u}_3[b], 0, \textbf{u}_3[q], 2\textbf{u}_3[q], 3\textbf{u}_3[q])$,
$\textbf{f}_4=(\textbf{u}_4[b], \textbf{u}_4[b] | 0, 0, 0, \textbf{u}_4[q],
\textbf{u}_4[q])$ and $\textbf{f}_5=(0, \textbf{u}_5[b] | 0, 0, 0, 0,
\textbf{u}_5[q])$ be representative vectors of the five row submatrices of
$\G_{BA}$.

\medskip

It is easy to see that the resulting vector $\textbf{u}$ after making any linear
combination between vectors in $\textbf{f}_1$ has minimum Lee weight $w_L(\textbf{u})=4d_\mathcal{X}$. We will now prove that any other linear combination between vectors in $\textbf{f}_1, \textbf{f}_2, \textbf{f}_3, \textbf{f}_4$ and $\textbf{f}_5$ gives a vector $\textbf{u}$ of minimum Lee weight $w_L(\textbf{u}) \geq 4d_\mathcal{X}$.
 
\medskip

Let us look at combinations between vectors in $\textbf{f}_2$. Considering the
$6$th and the $7$th column submatrices we have that $w_L(\textbf{u}_2[q]+\bar{\textbf{u}}_2[q]) +
w_L(3\textbf{u}_2[q]+ 3\bar{\textbf{u}}_2[q]) = w_L(2\textbf{u}_2[q] +
2\bar{\textbf{u}}_2[q])$, where $\bar{\textbf{u}}_2 \in
(\mathcal{Y}[b_2],\mathcal{Y}'[q_2])$. Let $\textbf{w}_1$ and
$\textbf{w}_2$ be two different binary vectors of the same length, such that
the Hamming distance between them is $d_b$. If we consider their binary zeros
and ones as zeros and ones in $\Z_4$, it is easy to see that the Lee distance
$d_q$ between any linear combination over $\Z_4$ of these vectors will be $d_q
\geq d_b$. Therefore, when we make linear combinations between any two column
rows of $\G_{BA}$, we have a lower bound on the minimum distance between them.
Since we can think of every vector in $\textbf{f}_2$ as if they were two vectors
in $\mathcal{Y}[bq_2]$, then vector $\textbf{u}$ in this case has
minimum Lee weight $w_L(\textbf{u}) \geq 2d_{\mathcal{Y}[bq_2]} \geq 2d_{\mathcal{Y}} = 4d_{\mathcal{X}}$.

Any linear combination between vectors in $\textbf{f}_3$, and also for those in $\textbf{f}_4$, yields a vector $\textbf{u}$ with minimum Lee weight $w_L(\textbf{u}) \geq 2d_{\mathcal{Y}[bq_4]} \geq 2d_\mathcal{Y} = 4d_\mathcal{X}$. 
Finally, linear combinations between vectors in $\textbf{f}_5$ yield a vector $\textbf{u}$ of minimum Lee weight $w_L(\textbf{u}) \geq d_\mathcal{Z} = 4d_\mathcal{X}$. 

\medskip 

Since $\mathcal{Y}[bq_4]$ $\subset$ $\mathcal{X}$ and $\mathcal{Z}$
$\subset$ $\mathcal{X}$, we have that $d(\textbf{u}_1,\textbf{u}_3)$, $d(\textbf{u}_1,\textbf{u}_4)$ and
$d(\textbf{u}_1,\textbf{u}_5)$ are greater than or equal to $d_\mathcal{X}$. Therefore, any linear combination between vectors
in $\textbf{f}_1$ and vectors in $\textbf{f}_3$, $\textbf{f}_4$  or $\textbf{f}_5$, will yield a vector $\textbf{u}$ of minimum Lee weight $w_L(\textbf{u}) \geq 4d_\mathcal{X}$.

If $\textbf{u}$ is the resulting vector of combining vectors in $\textbf{f}_3$
and vectors in $\textbf{f}_4$, its minimum Lee weight is clearly $w_L(\textbf{u})
\geq 2d_{\mathcal{Y}[bq_4]} \geq 2d_\mathcal{Y} = 4d_\mathcal{X}$.

Since $\mathcal{Z} \subset \mathcal{Y}$ and $\mathcal{Y}[bq_4] \subset
\mathcal{Y}$, we have $d(\textbf{u}_3,\textbf{u}_5)\geq d_{\mathcal{Y}}$ and
$d(\textbf{u}_4,\textbf{u}_5)\geq d_{\mathcal{Y}}$. Therefore, any linear
combination between vectors in $\textbf{f}_3$ and vectors in $\textbf{f}_5$, or between vectors in $\textbf{f}_4$ and vectors in
$\textbf{f}_5$ has minimum Lee weight $w_L(\textbf{u}) \geq 2d_\mathcal{Y} =
4d_\mathcal{X}$.

\medskip

It only remains to prove the minimum Lee weight of combinations between vectors
one of which is in $\textbf{f}_2$. Let $\textbf{u}$ be the resulting vector of making linear combinations between vectors in $\textbf{f}_1$ and vectors in $\textbf{f}_2$.

In terms of minimum distance, the third column submatrices of $\textbf{f}_1$ and
$\textbf{f}_2$ can be considered as a concatenation of two binary submatrices.
It is also easy to see that the minimum Lee distance between the quaternary
vectors $2\textbf{u}_1[b]$ and $\textbf{u}_2[b]$ is bigger than or equal to the minimum Hamming distance
between the binary vectors $(\textbf{u}_1[b],\textbf{u}_1[b])$ and
$(0,\textbf{u}_2[b])$, respectively. Therefore, since
$w_L(\textbf{u}_1[b]) + w_L(\textbf{u}_1[b] + \textbf{u}_2[b]) \geq
w_L(\textbf{u}_2[b])$ and $w_L(\textbf{u}_1[q] + \textbf{u}_2[q]) +
w_L(\textbf{u}_1[q] + 3\textbf{u}_2[q]) \geq w_L(2\textbf{u}_2[q])$ by the
triangle inequality, we have $w_L(\textbf{u}) \geq d_{\mathcal{Y}} +
2d_{\mathcal{X}} = 4d_{\mathcal{X}}$.

\bigskip

Let us now focus on combinations between vectors in $\textbf{f}_2$ and vectors in
$\textbf{f}_3$. Note that $w_L(\textbf{u}_2[q] + 2\textbf{u}_3[q]) +
w_L(3\textbf{u}_2[q] + 3\textbf{u}_3[q]) \geq w_L(2\textbf{u}_2[q] +
\textbf{u}_3[q])$ by the triangle inequality, and therefore the minimum Lee
weight of $\textbf{u}$ is $w_L(\textbf{u}) \geq 2d_{\mathcal{Y}} =
4d_{\mathcal{X}}$ because $w_L(\textbf{u}_2[b] + \textbf{u}_3[b]) +
w_L(2\textbf{u}_2[q] + \textbf{u}_3[q]) \geq d_{\mathcal{Y}}$.

\bigskip

If $\textbf{u}$ is now the resulting vector of making linear combinations between
vectors in $\textbf{f}_2$ and vectors in $\textbf{f}_4$, note that
$w_L(\textbf{u}_4[b]) + w_L(\textbf{u}_2[b] + \textbf{u}_4[b]) \geq
w_L(\textbf{u}_2[b])$ and $w_L(\textbf{u}_2[q]+\textbf{u}_4[q]) +
w_L(3\textbf{u}_2[q] + \textbf{u}_4[q]) \geq w_L(2\textbf{u}_2[q])$. Hence,
$w_L(\textbf{u}) \geq 2d_{\mathcal{Y}} = 4d_{\mathcal{X}}$.

\bigskip

Finally, let us look at combinations between vectors in $\textbf{f}_2$ and
vectors in $\textbf{f}_5$. Note that $w_L(\textbf{u}_2[q]) + w_L(3\textbf{u}_2[q]
+ \textbf{u}_5[q]) \geq w_L(2\textbf{u}_2[q] + \textbf{u}_5[q])$. Since
$\mathcal{Z} \subset \mathcal{Y}$ and $\mathcal{Y}[bq_2] \subset \mathcal{Y}$, we
have $d(\textbf{u}_3,\textbf{u}_5) \geq d_{\mathcal{Y}}$ and therefore
$w_L(\textbf{u}) \geq 2d_{\mathcal{Y}} = 4d_{\mathcal{X}}$.

Since all combinations of vectors in $\textbf{f}_1$, $\textbf{f}_2$,
$\textbf{f}_3$, $\textbf{f}_4$ and $\textbf{f}_5$ yield vectors $\textbf{v}$ of
minimum Lee weight $w_L(\textbf{u}) \geq 4d_{\mathcal{X}}$, we can conclude
that the minimum distance of any code generated by the BA-Plotkin construction
from Definition \ref{BAPlot} is $4d_{\mathcal{X}}$.

\end{proof}

\begin{lemm} \label{lemm:2u}
Let $\langle \textbf{u},\textbf{v} \rangle$ be a group of type
$2^14^1$ generated by $\textbf{u}$ and $\textbf{v}$, where
$\textbf{u}$,$\textbf{v}$ $\in$ $\Z_2^{\alpha}\times \Z_4^{\beta}$ are vectors of order two and four,
respectively. Then, $\textbf{u} \neq 2\textbf{v}$.
\end{lemm}

\begin{prop}\label{BA-PlotkinConsttype}
Code $\mathcal{C}$ generated by the BA-Plotkin
construction from Definition~\ref{BAPlot} is a $\zz$-additive code of type
$(2\alpha,\alpha+4\beta;\gamma,\delta)$ where $\gamma=\gamma_{\mathcal{X}}+\gamma_{\mathcal{Z}}$, $\delta=\delta_{\mathcal{X}}+\gamma_{\mathcal{Y}}+2\delta_{\mathcal{Y}}+\delta_{\mathcal{Z}}$, binary length $n=4\alpha+8\beta$ and size $2^{\gamma+2\delta}$.
\end{prop}

\begin{proof}

From the generator matrix $\G_{BA}$ of $\mathcal{C}$ it is straightforward to see that words in code $\mathcal{C}$ have $2\alpha$ binary and $\alpha+4\beta$ quaternary coordinates.

Note that the first row submatrix of $\G_{BA}$ consists of
$\gamma_\mathcal{X}$ order two and $\delta_\mathcal{X}$ order four row vectors; the second row submatrix has
$\gamma_\mathcal{Y}$ order four row vectors; the third and the fourth have
$2\delta_\mathcal{Y}$ order four row vectors altogether; and
the fifth row submatrix has $\gamma_\mathcal{Z}$ order two and
$\delta_\mathcal{Z}$ order four row vectors. 

Let $\mathcal{Y}[bq_2]$ and
$\mathcal{Y}[bq_4]$ be the codes generated by $\G_\mathcal{Y}[bq_2]$ and
$\G_\mathcal{Y}[bq_4]$, respectively.

If $\mathcal{Y}$ contained any vector $\textbf{u}$ $\in$ $\mathcal{Y}[bq_2]$
such that $\textbf{u}=2\textbf{v}$, where $\textbf{v}$ $\in$
$\mathcal{Y}[bq_4]$, then the group generated by vectors in the second and in
the fourth submatrices of $\G_{BA}$ would not be of type
$4^{\gamma_{\mathcal{Y}}+\delta_{\mathcal{Y}}}$, as we would expect. However,
since code $\mathcal{Y}$ is of type $(\alpha,\beta;\gamma_\mathcal{Y},\delta_\mathcal{Y})$, by
Lemma~\ref{lemm:2u} we know the situation just described cannot happen.
It is also easy to see that all five row submatrices from $\G_{BA}$ are independent to each other.
Therefore, the type of $\mathcal{C}$ is, indeed,
$(2\alpha,\alpha+4\beta;\gamma_{\mathcal{X}}+\gamma_{\mathcal{Z}},\delta_{\mathcal{X}}+\gamma_{\mathcal{Y}}+2\delta_{\mathcal{Y}}+\delta_{\mathcal{Z}})$.

\end{proof}

\begin{prop}\label{BA-PlotkinConstincl}
Let $\mathcal{W}$, $\mathcal{X}$, $\mathcal{Y}$ and $\mathcal{Z}$ be four
$\zz$-additive codes of 
%minimum distances $d_{\mathcal{W}}$, $d_{\mathcal{X}}$, $d_{\mathcal{Y}}$ and $d_{\mathcal{Z}}$ and 
generator matrices $\G_\mathcal{W}$, $\G_\mathcal{X}$, $\G_\mathcal{Y}$ and
$\G_\mathcal{Z}$, respectively, such that 
%$d_\mathcal{Z}=2d_\mathcal{Y}=4d_\mathcal{X}=8d_\mathcal{W}$, and
$\G_\mathcal{Z} \subset \G_\mathcal{Y} \subset \G_\mathcal{X} \subset \G_\mathcal{W}$.

The generator matrix obtained after applying the BA-Plotkin construction to
codes $\mathcal{W}$, $\mathcal{X}$ and $\mathcal{Y}$ contains the one
obtained after applying the same construction to codes
$\mathcal{X}$, $\mathcal{Y}$ and $\mathcal{Z}$.
\end{prop}

\begin{proof}
Straightforward, considering the inclusions $\G_\mathcal{Z}
\subset \G_\mathcal{Y} \subset \G_\mathcal{X} \subset \G_\mathcal{W}$.
\end{proof}

\section{Additive Reed-Muller codes}\label{sec3}

We will refer to $\zz$-additive Reed-Muller codes as $\ARM$.
Just as there is only one $RM$ family in the binary case, in the $\zz$-additive case there are $\lfloor \frac{m+2}{2} \rfloor$ families for each value of $m$. Each one of these families will contain any of the $\lfloor \frac{m+2}{2} \rfloor$ non-isomorphic $\zz$-linear extended perfect codes which are known to exist for any $m$ (see~\cite{BoRi99}).\\
We will identify each family ${\ARM}_s(r,m)$ by a subindex $s\in \{ 0,\ldots,\lfloor \frac{m}{2} \rfloor \}$.

\subsection{The families of $\ARM(r,1)$ and $\ARM(r,2)$ codes}\label{ARM2}

We start by considering the case $m=1$, that is the case of codes of
binary length $n=2^1$. The $\zz$-additive Reed-Muller code
$\ARM(0,1)$ is the repetition code, of type $(2,0;1,0)$, which has only one nonzero codeword
(the vector with only two binary coordinates of value $1$). The code $\ARM(1,1)$ is
the whole space $\Z_2^2$, thus a $\zz$-additive code of type
$(2,0;2,0)$. Both codes $\ARM(0,1)$ and $\ARM(1,1)$ are
binary codes with the same parameters and properties as the corresponding binary
$RM(r,1)$ codes (see \cite{Mac}). We will refer to them as $\ARM_0(0,1)$ and $\ARM_0(1,1)$,
respectively.

The generator matrix of $\ARM_0(0,1)$ is $\G_0(0,1)=\left(\begin{array}{cc}1 & 1\\
\end{array}\right)$ and the generator matrix of $\ARM_0(1,1)$ is
$\G_0(1,1)=\left(\begin{array}{cc}1 & 1 \\ 0 & 1\\
\end{array}\right)$.

For $m=2$ we have two families, $s=0$ and $s=1$, of additive Reed-Muller codes of binary length $n=2^2$.
The family $\ARM_0(r,2)$ consists of binary codes obtained
by applying the Plotkin construction defined in Proposition~\ref{PlotConst} to the family $\ARM_0(r,1)$. For $s=1$, we define $\ARM_1(0,2)$, $\ARM_1(1,2)$ and $\ARM_1(2,2)$ as the codes with generator
matrices $\G_1(0,2)=\left(\begin{array}{cc|c}1 & 1 & 2\\
\end{array}\right)$,
$\G_1(1,2)=\left(\begin{array}{cc|c}1 & 1 & 2 \\ \hline 0 & 1 & 1\\
\end{array}\right)$ and $\G_1(2,2)=\left(\begin{array}{cc|c}1 & 1 & 2 \\ 0 & 1 & 0 \\ \hline 0 & 1 & 1\\
\end{array}\right)$, respectively.

\subsection{Plotkin and BA-Plotkin constructions}\label{construccions}

Take the family $\ARM_s$ and let $\ARM_s(r,m-1)$, $\ARM_s(r-1,m-1)$ and $\ARM_s(r-2,m-1)$,
 $0\leq s\leq \lfloor \frac{m-1}{2}\rfloor $, be three consecutive codes
with parameters ($\alpha,\beta;\gamma',\delta'$), ($\alpha,\beta;\gamma'',\delta''$) and ($\alpha,\beta;\gamma''',\delta'''$);
binary length $n=2^{m-1}$; minimum distances $2^{m-r-1}$, $2^{m-r}$ and $2^{m-r+1}$; and generator matrices $\G_{s}(r,m-1)$, $\G_{s}(r-1,m-1)$ and $\G_{s}(r-2,m-1)$, respectively.
By using Proposition~\ref{PlotConst} and Proposition~\ref{BA-PlotkinConst} we can prove the following results:

\begin{theo}\label{plotkin}
For any $r$ and $m\geq 2$, $0< r< m$, code $\ARM_s(r,m)$ obtained
by applying the Plotkin construction from Definition~\ref{PlotDefi} on codes $\ARM_s(r,m-1)$ and $\ARM_s(r-1,m-1)$ is
a $\zz$-additive code of type ($2\alpha$, $2\beta$; $\gamma$, $\delta$), where
$\gamma=\gamma'+\gamma''$ and $\delta=\delta'+\delta''$; binary
length $n=2^{m}$; size $2^k$ codewords, where $k= \dd
\sum_{i=0}^r\binom{m}{i}$; minimum distance $2^{m-r}$ and
$\ARM_s(r-1,m) \subset \ARM_s(r,m)$.

We consider $\ARM_s(0,m)$ to be the repetition code with only one nonzero codeword (the vector with $2\alpha$ ones and $2\beta$ twos) and $\ARM_s(m,m)$ be the whole space $\Z_2^{2\alpha}\times\Z_4^{2\beta}$.
\end{theo}

\begin{theo}\label{BA-plotkin}
For any $r$ and $m\geq 3$, $0< r< m$, $s>0$,
use the BA-Plotkin construction from Definition~\ref{BAPlot}, where generator matrices $\G_{\mathcal{X}}$, $\G_{\mathcal{Y}}$, $\G_{\mathcal{Z}}$ stand for $\G_{s}(r,m-1)$, $\G_{s}(r-1,m-1)$ and $\G_{s}(r-2,m-1)$, respectively, to obtain a new $\zz$-additive 
$\ARM_{s+1}(r,m+1)$ code of type ($2\alpha$, $\alpha+4\beta$; $\gamma$, $\delta$),
where $\gamma=\gamma'+\gamma'''$, $\delta=\delta'+\gamma''+2\delta''+\delta'''$; binary length
$n=2^{m+1}$; $2^k$ codewords, where $k=~ \dd
\sum_{i=0}^r\binom{m+1}{i}$, minimum distance $2^{m-r+1}$ and, moreover,
$\ARM_{s+1}(r-~1,m+1)~\subset~\ARM_{s+1}(r,m+1)$.
\end{theo}

To be coherent with all notations, code
$\ARM_{s+1}(-1,m+1)$ is defined as the all zero codeword code, code $\ARM_{s+1}(0,m+1)$ is defined as the repetition code
with only one nonzero codeword (the vector with $2\alpha$ ones and $\alpha+4\beta$ twos), whereas codes $\ARM_{s+1}(m,m+1)$ and
$\ARM_{s+1}(m+1,m+1)$ are defined as the even Lee weight code and the
whole space $\Z_2^{2\alpha}\times\Z_4^{\alpha+4\beta}$, respectively.

Using both Theorem~\ref{plotkin} and Theorem~\ref{BA-plotkin} we can now
construct all $\ARM_{s}(r,m)$ codes for $m>2$. Once applied the Gray map, all these codes give rise to binary codes with the
same parameters and properties as the $RM$ codes. Moreover, when $m = 2$ or $m = 3$, they also have the same
codewords.

\end{document}